\documentstyle[emulateapj]{article}
\def\beq{\begin{equation}}
\def\eeq{\end{equation}}
\def\bey{\begin{eqnarray}}
\def\eey{\end{eqnarray}}
\def\kms{{\rm \,km\,s^{-1}}}

\def\kpc{\,{\rm {kpc}}}

\def\chisq{\chi^2}

\def\tE{t_{\rm E}}
\def\rEt{\tilde{r}_{\rm E}}
\def\rEp{r_{\rm E,p}}

\def\thetaE{\theta_{\rm E}}
\def\thetastar{\theta_\star}
\def\tv{\tilde{v}}
\def\u0{u_0}
\def\tp{t_{\rm p}}

\def\Ds{D_{\rm s}}
\def\Dd{D_{\rm d}}
\def\mI0{m_{I,0}}
\def\ustar{u_\star}
\def\phig{\phi_\gamma}
\def\eps{\epsilon}
\def\rad{\thinspace{\rm rad}}
\def\nhat{\hat{n}}

\def\xlens{x_{\rm L}}
\def\ylens{y_{\rm L}}
\def\zlens{z_{\rm L}}
\def\xearth{x_\oplus}
\def\yearth{y_\oplus}

\def\mas{\mu{\rm as}}
\def\zearth{z_\oplus}
\def\Drel{D_{\rm rel}}

\def\murel{\mu_{\rm rel}}
\def\pirel{\pi_{\rm rel}}
\def\deltar{\delta{\vec{r}}}

\def\Is{I_{\rm s}}
\newcommand{\zdot}{\makebox[0pt][l]{.}}
\newcommand{\up}[1]{\ifmmode^{\rm #1}\else$^{\rm #1}$\fi}

\newcommand{\uph}{\up{h}}
\newcommand{\upm}{\up{m}}
\newcommand{\ups}{\up{s}}
\newcommand{\arcd}{\ifmmode^{\circ}\else$^{\circ}$\fi}
\newcommand{\arcm}{\ifmmode{'}\else$'$\fi}
\newcommand{\arcs}{\ifmmode{''}\else$''$\fi}

\begin{document}

\def\thefootnote{\fnsymbol{footnote}}

\title{Optical Gravitational Lensing Experiment.\\
OGLE-2000-BUL-43: A Spectacular Ongoing Parallax Microlensing Event.\\
Difference Image Analysis.
\footnotemark}

\author{I. Soszy\'nski$^{1,2}$, K. \.Zebru\'n$^{1,2}$, P.R. Wo\'zniak$^{2}$,
S. Mao$^{3}$, A. Udalski$^{1}$, M. Szyma\'nski$^{1}$, M.~Kubiak$^{1}$,
G. Pietrzy\'nski$^{1,4}$, O. Szewczyk$^{1}$, \L. Wyrzykowski$^{1}$.}

\affil{$^{1}$Warsaw University Observatory, Al. Ujazdowskie 4,
00-478 Warszawa, Poland}
\affil{e-mail: soszynsk,zebrun,udalski,msz,mk,pietrzyn,szewczyk,wyrzykow@astrouw.edu.pl}
\affil{$^{2}$Princeton University Observatory, Princeton, NJ 08544--1001, USA}
\affil{e-mail: soszynsk,zebrun,wozniak@astro.princeton.edu}
\affil{$^{3}$ Univ. of Manchester, Jodrell Bank Observatory, Macclesfield,
Cheshire SK11 9DL, UK} 
\affil{e-mail: smao@jb.man.ac.uk}
\affil{$^{4}$ Universidad de Concepci{\'o}n, Departamento de Fisica,
Casilla 160--C, Concepci{\'o}n, Chile}

\bigskip

\begin{abstract}

We present the photometry and theoretical models for a Galactic bulge
microlensing event OGLE-2000-BUL-43. The event is very bright with
$I=13.54$ mag, and has a very long time scale, $\tE=156$ days. The long
time scale and its light curve deviation from the standard shape strongly
suggest that it may be affected by the parallax effect. We show that
OGLE-2000-BUL-43 is the first discovered microlensing event, in which the
parallax distortion is observed over a period of 2 years. Difference Image
Analysis (DIA) using the PSF matching algorithm of Alard \& Lupton enabled
photometry accurate to 0.5\%. All photometry obtained with DIA is available
electronically.  Our analysis indicates that the viewing condition from a
location near Jupiter will be optimal
and can lead to magnifications $\sim 50$ around January 31, 2001.
These features offer a
great promise for resolving the source (a K giant)
and breaking the degeneracy between the lens parameters including
the mass of the lens, if the event is observed with the imaging camera on
the Cassini space probe.

\end{abstract}


\footnotetext{Based on observations obtained with the 1.3 m Warsaw
Telescope at the Las Campanas Observatory of the Carnegie
Institution of Washington.}

\section{Introduction}

Gravitational microlensing was originally proposed as a method of detecting
compact dark matter objects in the Galactic halo (Paczy\'nski
1986). However, it also turned out to be an extremely useful method to
study Galactic structure, mass functions of stars and potentially extra-solar
planetary systems (for a review, see Paczy\'nski 1996). 
Most microlensing events are well described by the standard light curve
(e.g., Paczy\'nski 1986). Unfortunately, from these light curves, one
can derive only a single physical constraint, namely the Einstein radius
crossing time, which involves the lens mass, various distance measures
and relative velocity (see \S 4). This degeneracy means that the lens
properties cannot be uniquely inferred. Therefore any further information
on the lens configuration is of great importance. Microlensing events
that exhibit parallax effects provide this type of information. Such
effects can occur when the event is observed simultaneously from
two different positions in the Solar system (Refsdal 1966) or when the
event lasts long enough that the Earth's motion can no longer be
approximated as rectilinear during the event (Gould 1992).  Both of these
effects will be directly relevant to the current paper. The first
parallax microlensing event was reported 
by the MACHO collaboration toward the Galactic bulge (Alcock et al. 1995),
and the second case (toward Carina) was discovered by the OGLE collaboration
and reported in Mao (1999). Additional parallax microlensing
candidates have been presented in a conference proceeding
(Bennett et al.\ 1997). In this paper, we report a new parallax
microlensing event, OGLE-2000-BUL-43. This bulge event was discovered well
ahead of the peak by the Early Warning System (Udalski et al. 1994), and
attracted attention due to its extreme brightness and very long time
scale.

The unusually long duration of the event ($t_{\rm E}\sim 156\,$days)
combined with the extremely small velocity of the magnification
pattern on the plane of the observer ($\tilde v \sim 40\,\kms$, i.e.,
hardly faster than the motion of the Earth), imply that the parallax
effect is not only detectable, but measurable very precisely.  To make 
the most of this possibility, we employ difference image analysis (DIA,
Wo\'zniak 2000) to optimize the photometry (\S 2).

The parallax measurement that we present here yields not only the
size of the Einstein radius projected onto the observer plane
($\tilde r_{\rm E} \approx 3.62\,$AU) but also the direction of lens-source
relative motion in the heliocentric coordinate system.
 By combining these two, we can predict the light curve seen
by any observer in the solar system as a function of time. 
In particular, we predict that as seen from the Cassini spacecraft around
January 31, 2001,  
the lens and source will have an extraordinarily close separation
and hence the source will be highly magnified.  Unless the lens turns out to be
very massive ($M\ga 0.8\,M_\odot$) and close ($\Dd\la 1\,$kpc), such
a separation would permit resolution of the source and hence
measurement of the the angular Einstein radius, $\thetaE$
(Alcock et al.\ 1997, 2000; Albrow et al.\ 1999, 2000, 2001;
Afonso et al.\ 2000).  Gould (1992) showed that by combining measurements
of $\tilde r_{\rm E}$, $\thetaE$ and the Einstein radius crossing time
($\tE$), one could obtain a complete solution of the event,
\bey
M & = & {c^2 \over 4G} \rEt \thetaE, \label{M} \\
\Drel & = & {\rEt \over \thetaE} \equiv {{\rm AU}
\over \pirel}, \label{Drel} \\
\murel & = & {\thetaE \over \tE}, \label{murel}
\eey
where $\Drel=\Dd\Ds/(\Ds-\Dd)$, $\pirel={\rm AU}/\Dd-{\rm AU}/\Ds$ is the
lens-source relative parallax, $\murel$ is the lens-source
relative proper motion, and $\Dd$ and $\Ds$ are the
distances to the lens and source, respectively. See also Gould (2000).
Since the source is quite bright even at baseline $(I=13.54, V=15.65)$
it should be easily measurable by the Cassini probe. Cassini photometry would
therefore very likely yield the first mass measurement of a microlensing
event.

The outline of the paper is as follows. In \S2 we describe observations,
in \S3 we describe our photometric reduction method, \S4 contains the
details of model fitting and predicted viewing conditions, in \S5 we
describe potential scientific returns of Cassini observations and
finally in \S6, we briefly summarize and discuss our results.

\section{Observations} 

All observations presented in this paper were carried out during the 
second phase of the OGLE experiment with the 1.3-m Warsaw 
telescope at the Las Campanas Observatory, Chile, which is 
operated by the Carnegie Institution of  Washington. The telescope 
was equipped with the ``first generation'' camera with a SITe 
${2048\times2048}$ CCD detector working in the drift-scan mode. The  
pixel size was 24~$\mu$m giving the scale of 0.417$\arcsec$ per
pixel. Observations 
of  the Galactic bulge fields were performed in the ``medium'' reading mode 
of the CCD detector with the  gain 7.1~e$^-$/ADU and readout noise about 
6.3~e$^-$. Details of the instrumentation setup can be found in Udalski, 
Kubiak \& Szyma{\'n}ski (1997). 

The OGLE-2000-BUL-43 event was detected by the OGLE Early Warning System
(Udalski et al. 1994) in mid-2000. Equatorial coordinates of the event for
2000.0 epoch are:
$\alpha_{2000}=18\uph08\upm43\zdot\ups04$,
$\delta_{2000}=-32\arcd24\arcm39\zdot\arcs5$, ecliptic
coordinates are: $\lambda=271\zdot\arcd863$, $\beta=-8\zdot\arcd986$ and
Galactic coordinates are $l=359\zdot\arcd467$, $b=-6\zdot\arcd036$.
Figure~1 is a finding chart
showing the $120\arcsec\times 120\arcsec$ region centered on the event. 
Observations of this field started in March 1997, and continued until
November~22, 2000. The bulge observing season usually ends at the beginning
of November, therefore the latest observations of OGLE-2000-BUL-43 were made
in difficult conditions with the object setting shortly after the sunset, when
the sky is still quite bright. Fortunately the source was bright enough so that
poor seeing and high backgrounds were not a significant problem in the DIA
analysis.

The majority of the OGLE-II frames are taken in the $I$-band. For the BUL\_SC7
field 330 $I$-band and 9 $V$-band observations were collected. Udalski et
al. (2000) gives full details of the standard OGLE observing techniques
and the DoPhot photometry is available from OGLE web site at
{\it http://www.astrouw.edu.pl/$\sim$ogle/ogle2/ews/ews.html}.

\section{Photometry}

Our analysis includes all $I$-band observations of the BUL\_SC7 field. 
We used the DIA technique to obtain the light curves
of the OGLE-2000-BUL-43 event. Our method is based on the recently
developed optimal PSF matching algorithm (Alard \& Lupton 1998; Alard
2000). Unlike other methods that use divisions in Fourier space (Crotts
1992, Phillips \& Davis 1995, Tomaney \& Crotts 1996, Reiss et al. 1998,
Alcock et al. 1999), the Alard \& Lupton method operates directly in real
space. Additionally it is not required to know the PSF of each image to
determine the convolution kernel. Wo\'zniak (2000) tested the method on
large samples and showed that the error distribution is Gaussian to better
than 1\%. Compared to the standard DoPhot photometry (Schechter, Mateo, \&
Saha 1993), the scatter was always improved by a factor of 2--3 and frames
taken in even the worst seeing conditions gave good photometric points.

Our DIA software handles PSF variations in drift-scan images by polynomial
fits.
Even then it is required that the frames are subdivided into 512$\times$128
pixel strips because PSF variability along the direction of the scan is much
faster than across the frame. The object of interest turned out to be not too
far from the center of one of the subframes selected automatically, therefore
we basically adopted the standard pipeline output for that piece of the sky
without the need to run the software on the full format.
Minor modifications included more careful preparation of the reference image
and calibration of the counts in terms of standard magnitude system. 

First, from the full data set for the BUL\_SC7 field we selected 20 frames
with the best seeing, small shifts relative to OGLE template and low background level. More
weight was assigned to the PSF shape and quality of telescope tracking
in the analyzed region during the selection process. These frames were co-added
to create a reference frame for all subsequent subtractions. Preparation
of the reference image was absolutely critical for the quality of
the final results.

Next we ran the DIA pipeline for all of our data to retrieve the AC signal (variable
part of the flux) of our lensed star. The software rejected only 9 frames due
to very bad observing conditions or very large shifts in respect to the
reference image. Our final light curve
contains 321 observations. To calibrate the result on the magnitude scale
we ran DoPhot on the reference image. The magnitude zero point
($I=13.54, V=15.65$) was obtained by comparing our DoPhot photometry
with the OGLE database. 

The DIA light curve is shown in Figure~2. The scatter in the photometry is
0.5\% and is dominated by systematics due to atmospheric turbulence and PSF
variations. The individual error bars returned by the automated massive
photometry pipeline (Wo\'zniak 2000) proved to be over-estimated
when compared to the scatter around the best fit model
(\S4). Most likely this is a combined result of individual care during
data processing for OGLE-2000-BUL-43 and relatively low density of stars
in the BUL\_SC7 field. The errors were re-calibrated so as to
enforce $\chi^2$ per degree of freedom to be unity in the best-fit model
with parallax (see \S4 and Table 1).

We would like to stress the fact that it is the accuracy achieved here with
the DIA method which enabled a detailed study of the lens parameters.
Figure~3 presents the distribution of residuals with respect to the model
(see \S4) for measurements with the DIA pipeline. Maximal differences
between the classical single point microlensing model and the parallax fit
are indicated by dashed vertical lines. One notices that the scatter of the
photometry is small enough to analyse the parallax effect.
Additionally our data set contains 82 more points than the OGLE
EWS light curve. The difference is because the lowest grade frames are
rejected in the standard DoPhot analysis. The DIA photometry data file is
available from the OGLE anonymous FTP server: {\it
ftp://sirius.astrouw.edu.pl/ogle/ogle2/BUL-43/bul43.dat.gz}.

In Figure~4 we present the Color-Magnitude Diagram for the BUL\_SC7 field. The position of the lensed star
(marked by a cross) suggests that the source is a K giant. For later
studies of the finite source size effect (\S 5), we would
like to estimate the angular diameter of the star. In order to do this,
we first need to estimate the dereddened color and magnitude of
the star. For this purpose, we use the red-clump giants
that have well calibrated dereddened colors and magnitudes. We adopt the
average color and magnitude of red-clump giants in Baade's window
from the previous studies (Paczy\'nski et al. 1999),
\beq
(V-I)_{\rm RC, 0} = 1.11, ~~ I_{\rm RC, 0} = 14.37.
\eeq 
From Figure~4, the red-clump stars in the BUL\_SC7 field have
\beq
(V-I)_{\rm RC}=1.67 \pm 0.02, ~~
I_{\rm RC}=15.15 \pm 0.05. 
\eeq
Hence we have
\beq
E(V-I) = (V-I)_{\rm RC} - (V-I)_{\rm RC,0} = 0.57,
\eeq
and
\beq
A_I=I_{\rm RC}-I_{\rm RC,0}=0.78.
\eeq
Taking into account a blending parameter $f=0.91$ (see \S 4, Table 1) in
the $I$-band, we
obtain the magnitudes of the lensed star as $I=13.64, V=15.75$. Hence, 
the intrinsic color and $I$-band magnitude for OGLE-2000-BUL-43 are
\beq
(V-I)_0 = (V-I)-E(V-I) = 1.54, ~~ I_0= I-A_I = 12.86.
\eeq
Note that these values we derived are somewhat different from those of
Schlegel, Finkbeiner \& Davis (1998): $A_I=0.92$ and $E_{V-I}=0.61$.
Our smaller extinction value is consistent with
Stanek (1998) who argued that the Schlegel et al. (1998) map
over-estimates the extinction for $|b|<5^\circ$. Our estimate based on
the red-clump giants is also somewhat uncertain because of the
metallicity gradient that may exist between Baade's window ($b=-4^\circ$)
and the BUL\_SC7 field ($b=-6^\circ$). Fortunately, this uncertainty 
in reddening only affects the angular diameter estimate very
slightly because the surface brightness-color relation has a slope
similar to the slope of the reddening line (see below).

Using the dereddened color and magnitude,
we can estimate the angular stellar radius ($\thetastar$) using
the empirically determined relation between color and surface
brightness (van Belle 1999), independent of the source
distance. Transforming van Belle's relation given in 
$V$ vs. $V-K$ into $I$ vs. $V-I$ using the color-color relations of
Bessel \& Brett (1988), one obtains
\beq
\thetastar = 18.9 \mas\, \times 10^{(12.90-I_0)/5} \times [(V-I)_0-0.6]
\eeq
For our star, this gives $\thetastar=18.1\mas$.
Using the values from Schlegel et al. (1998), 
the $\thetastar$ value increases by about 2\%. Therefore the estimate
of the angular stellar radius is quite robust.

\section{Model}

We first fit the light curve with the standard single microlens
model which is sufficient to describe most microlensing events.
In this model, the (point) source, the lens and the observer all
move with constant spatial velocities. The standard form is given by
(e.g., Paczy\'nski 1986):
\beq \label{amp}
A(t) = {u^2+2 \over u \sqrt{u^2+4}},~~
u(t) \equiv \sqrt{\u0^2 + \tau(t)^2},
\eeq
where $\u0$ is the impact parameter (in units of the Einstein radius) and 
\beq \label{tau}
\tau(t) = {t-t_0 \over \tE}, ~~ \tE = {\thetaE \over \murel},
\eeq
with $t_0$ being the time of the closest approach (maximum
magnification), $\thetaE$ the angular Einstein radius, 
and $\tE$ the Einstein radius crossing time. The explicit forms
of the angular Einstein radius ($\thetaE$) and the projected Einstein
radius ($\rEt$) are
\beq \label{thetaE}
\thetaE = \sqrt{{4 G M \over c^2 \Drel}}, ~~
\rEt = \sqrt{{4 G M \Drel \over c^2}}
\eeq
where $M$ is again the lens mass and $\Drel$ is defined below equation (\ref{murel}).
For microlensing in the local group, $\thetaE$ is $\sim$ mas and $\rEt$
$\sim$ few AU.
Equations (\ref{amp}-\ref{thetaE}) show the well-known lens degeneracy, i.e.,
from a measured $\tE$, one can not infer the lens mass, distances and
kinematics uniquely even if the source distance is known.

To fit the $I$-band data with the standard model, we need a minimum of four
parameters, namely, 
$\u0, t_0, \tE, \Is$,
where $\Is$ is the unlensed $I$-band magnitude of the source.
The best-fit parameters (and their errors)
are found by minimizing the usual $\chisq$ using the MINUIT program in the
CERN library$\footnote{http://wwwinfo.cern.ch/asd/cernlib/}$
and are tabulated in Table 1 (model S). The resulting $\chisq$ is 9025.2 for 317 degrees of freedom. 
The large $\chisq$ indicates that the fit is unacceptable. This can also be
clearly seen in Figure~2, where we have plotted the predicted light curve
as the dotted line. The deviation is apparent in the 2000 observing season.
In fact, upon closer examination, the model over-predicts the
magnification in the 1999 season as well (see the bottom inset in Figure~2).
Since the Galactic bulge fields are very crowded, there could be some
blended light from a nearby unlensed source within the seeing disk of
the lensed source, or there could be some light from the lens itself. So 
in the model we can introduce a blending parameter, $f$, which we define as the fraction
of light contributed by the lensed source in the baseline ($f=1$ if there is no
blending). Note that blending is introduced in our adoption of
the magnitude zero point obtained by the DoPhot photometry; the DIA
method itself automatically subtracts out the blended light. The inclusion of the blending
parameter reduces the $\chisq$ to 2778.4 for 316 degrees of freedom.
This model requires a blending fraction $f=0.22$, which is implausible
considering the extreme brightness of the lensed star.
In any case, the $\chi^2$ is better but still far from acceptable. 
We show below that all these discrepancies can be removed by 
incorporating the parallax effect.

To account for the parallax effect, we need
to describe the Earth motion around the Sun. We adopt a heliocentric
coordinate system with the $z$-axis toward the Ecliptic north and the
$x$-axis from the Sun toward the Earth at the Vernal
Equinox\footnote{Another commonly used
heliocentric system (e.g., in the Astronomical
Almanac 2000) has the $x$-axis
opposite to our definition.}. The position of the Earth,
to the first order of the orbital eccentricity ($\epsilon \approx 
0.017$), is then (e.g., Dominik 1998 and references therein)
\bey
\nonumber\xearth(t) & = & A(t) \cos [\xi(t)-\phig],\\
\yearth(t) & = & A(t) \sin [\xi(t)-\phig],\\
\nonumber\zearth(t) & = & 0,
\eey
where 
\beq
A(t) = {\rm AU} \, (1-\eps \cos \Phi), ~~~
\xi(t) = \Phi +2\eps \sin\Phi
\eeq
with $\Phi=2\pi(t-\tp)/T$, $T=1\,{\rm yr}$, and
$\phig\approx 75\zdot\arcd98$ is the longitude difference between the
Perihelion ($\tp=1546.708$) and the Vernal Equinox
($t \equiv {\rm JD}-2450000=1623.816$) for J2000. The line of sight in the heliocentric
coordinate system is as usual described by
two angular polar coordinates $(\phi, \chi)$. These two angles are
related to the geocentric ecliptic coordinates $(\lambda, \beta)$ by
$\chi=\beta$, and $\phi=\pi+\lambda$. Again, for OGLE-2000-BUL-43,
$\beta=-8\zdot\arcd986$, and $\lambda=271\zdot\arcd863$ (see, e.g., Lang 1981
for conversions between different coordinate systems).

To describe the lens parallax effect, we find it more intuitive
to use the natural formalism as
advocated by Gould (2000), i.e., we project the usual lensing quantities
into the observer (and ecliptic) plane. The line of sight vector is given by
$\nhat = (\cos\chi\cos\phi, \cos\chi\sin\phi, \sin\chi)$ in the heliocentric
coordinate system. For a vector, $\vec{r}$, the component perpendicular to
the line of sight is given by
$\vec{r}_\perp=\vec{r}-(\vec{r}\cdot\nhat)\nhat$. For example,
the perpendicular component of the Earth position is
$\vec{r}_{\oplus,\perp}={\vec{r}_\oplus-(\vec{r}_\oplus \cdot\nhat)\nhat}$.
Thus, a circle in the lens plane ($\vec{r}_\perp^2=R^2$) 
is mapped into an ellipse in the ecliptic plane, which is given by,
\beq \label{project}
r = {R \over \sqrt{1- \cos^2\chi \cos^2(\Theta-\phi)}},
\eeq
where $\Theta$ is the polar angle in the ecliptic plane. The minor axis
and major axis for the ellipse are $R$ and $R/\sin\chi$, respectively. 

The lens trajectory is described by two parameters, the dimensionless impact
parameter, $\u0$, and the angle, $\psi$, between
the heliocentric ecliptic $x$-axis and the normal to the trajectory. Note that $\u0$
is now more appropriately the (dimensionless)
minimum distance between the Sun-source line and the
lens trajectory. For convenience, we define the Sun to be on
the left-hand side of the lens trajectory for $\u0>0$.
The lens position (in physical units) projected into the ecliptic plane,
$\vec{r}_{\rm L}=(\xlens, \ylens, 0)$, 
as a function of time, is given by
\bey
\nonumber\xlens & = & \u0\rEt\cos\psi - \tau \rEp(\psi) \sin\psi,\\
\ylens & = & \u0\rEt\sin\psi + \tau \rEp(\psi) \cos\psi,\\
\nonumber\zlens & = & 0,
\eey
where $\tau$ and $\rEt$ are defined in equations (\ref{tau}) and
(\ref{thetaE}), and $\rEp=\rEt/\sqrt{1-\cos^2\chi\sin^2(\pi/2+\psi-\phi)}$
is the Einstein radius projected into the ecliptic plane in
the direction of the lens trajectory. The expression of
$\rEp$ can be derived using equation (\ref{project}) 
with $\Theta=\pi/2+\psi$, where the factor
$\pi/2$ arises because $\psi$ is defined as
the angle between the normal to the trajectory and the $x$-axis. We denote
the vector from the lens position (projected into the ecliptic plane)
toward the Earth as $\deltar=\vec{r}_{\oplus}-\vec{r}_{\rm L}$.  The
component of $\deltar$ perpendicular to the line of sight is 
$\deltar_\perp = \deltar-(\deltar\cdot \nhat) \nhat$.
The magnification can then be calculated using equation (\ref{amp}) with
$u^2=(\deltar_\perp/\rEt)^2$.

In total, seven parameters ($\u0, t_0, \tE,\Is, \rEt, \psi, f$)
are needed to describe the parallax effect with blending.
These parameters are again found by minimizing $\chisq$.
In table 1, we list the best fit parameters (model P); for this model,
the $\chisq$ per degree of freedom is now unity due to our
rescaling of errors (see \S2). In particular, we find that
\beq
\rEt = (3.62\pm 0.16)\thinspace {\rm AU}, ~~\psi=(3.024\pm 0.005) \rad.
\eeq
The correlation coefficient between $\rEt$ and $\psi$ is $-0.088$.
The predicted light curve is shown in Figure~2 as the solid line. The model
fits the data points very well. Notice that the model requires a
marginal blending with $f=0.911\pm 0.056$. This is expected
since the source star is very bright, and it appears unlikely that any
additional source can contribute substantially to the total light. We
return to the degeneracy of solutions briefly in \S6.

Using equations (\ref{M}-\ref{Drel}), and $\rEt \approx
 3.62\,{\rm AU}$,  we obtain the lens mass as a function of the relative 
lens-source parallax
\beq \label{Mestimate}
M={c^2 \rEt^2 \over 4 G} \pirel 
=0.23 M_\odot \left({3.5\kpc \over \Dd}-{7\kpc \over \Ds}\right).
\eeq
So the lens is likely to be low-mass unless it is unusually close to us
($\Dd \sim 1\kpc$).
Combining $\rEt$ and $\tE$, we can also derive the projected velocity of the
lens,
\beq
\tv = \murel \Drel = {\rEt \over \tE} = (40\pm 2) {\kms}.
\eeq
The low projected velocity favors a disk-disk lensing event. For such
events, the observer, the lens and the source rotate about the Galactic
center with roughly the same velocity, and the relative motion is only
due to the small, $\sim 10\kms$, random velocities 
(see, e.g., Derue et al. 1999).
On the other hand, the chance for a bulge source (with its much larger
random velocity, $\sim 100\kms$) 
to have such a low projected velocity relative to
the lens (whether disk or bulge) is small. The low projected speed and
the long duration of this event imply that the Earth's motion induces
a large excursion in the Einstein ring, and this large deviation from
rectilinear motion makes an accurate parallax measurement possible, even
though the event has only barely reached its peak.

The accurate measurement of $\rEt$ and $\psi$ makes it possible to
predict the light curve that would be seen by a hypothetical observer
anywhere in the solar system. 
Figure~5 shows the illumination pattern on January 1.000, 2001 UT.
The two elliptical curves are iso-magnification contours for
$A=1.342$ and 4, respectively; the outer contour with $A=1.342$
corresponds to the Einstein `ring' in the ecliptic
plane. It appears as an ellipse in Figures~5 and 7 because the ecliptic
plane is not perpendicular to the source direction (cf. eq.~\ref{project}).
Various filled dots indicate the positions of the
source, Earth, Jupiter and Saturn on this date.
The open dots indicate the positions of the
source and the planets every half a year in the future.
From this figure, one can see that the inner contour nearly coincides with the
position of Jupiter on January 1, 2001, hence an observer close to Jupiter will see a
magnification of about 4; the magnification is even higher somewhat later.
The Cassini probe is currently approaching Jupiter,
for a fly-by acceleration on its way to Saturn, it is therefore
an ideal instrument to observe this event from space.
In the next section, we will discuss in some detail the
potential scientific returns of Cassini observations.

\section{Potential Scientific Returns of Cassini Observations}

In Figure~6 we show the light curve of OGLE-2000-BUL-43 for an 
observer near Jupiter, mimicking the fly-by observations from Cassini.
The light curve shows a spectacular peak at
JD$\approx$ 2451940.5 (January 31, 2001). Figure~7 illustrates the position of
Jupiter with respect to the illumination pattern. It clearly shows
that the lens and Jupiter will come very close together
and hence one will see a very high magnification around that time.

When the physical impact parameter 
is comparable to the stellar radius, microlensing light curves
are substantially modified by the finite source size effect
(Gould 1994; Nemiroff \& Wickramasinghe 1994; Witt \& Mao 1994). More
precisely, when
\beq
\u0 \la \ustar \equiv {\thetastar \over \thetaE} =
{c^2 \over 4 GM} \rEt \thetastar =  {0.008 \over M/M_\odot},
\eeq
then finite source size effects will be significant and it becomes
feasible to measure $\thetaE$, hence providing one more constraint
on the lens parameters. Our best-fit model has a minimum impact
parameter (in the lens plane) $\u0 = 3.6\times 10^{-3}$, and so unless the lens is very close to us
and very massive (eq. \ref{Mestimate}), the finite source size 
will be resolved. The inset in Fig.~6 illustrates this effect
where we have adopted $\thetaE=0.47$\,mas.
The effect is quite dramatic. In comparison, the effect is negligible
for an observer on Earth. Note that the peak of the light curve only depends 
on $\ustar=\thetastar/\thetaE$. So the peak can be
higher if the angular Einstein radius is larger, and vice versa.

To plan space observations, it is important to estimate the errors
in the minimum impact parameter ($\u0$) and the peak time ($t_0$). We have
performed Monte Carlo simulations to estimate their uncertainties
(e.g., Press et al. 1992). We find that the 95\% confidence limits on
$u_0$ and $t_0$ are $10^{-4}<u_0<0.011$ and $1938.3<t_0<1941.3$,
respectively. It is therefore very likely that the magnification at
Jupiter will be very high. The peak time is accurate to about 3 days
while the finite source size effect lasts for about twenty days 
(see the inset in Figure~6). To detect this effect, it is crucial to 
have at least a few observations during the lens transit across the 
stellar surface (Peng 1995).
If the finite source size effect is indeed observed by
Cassini, then we can measure $\thetaE$, and this will lead to, for the
first time, a complete solution of the lens parameters, including the
lens mass, the relative lens-source parallax and proper motions (see
introduction). We again emphasize 
that the determination of mass is independent of the source distance
if $\thetaE$ is measured (cf. eq.~\ref{M}).

\section{Summary and Discussion}

OGLE-2000-BUL-43 is the longest microlensing event observed by the OGLE
project. It is also the first event, in which the parallax effect is
observed over a 2 year period, making the association of the acceleration
term with the motion of the Earth unambiguous. Photometric accuracy at the 0.5\%
level enabled a detailed study of the event parameters partly removing
the degeneracy between the mass, velocity and distance. We conclude that
the lens is slow moving, and unless it is unusually close to us, the lens
mass is expected to be small.

The main aim of this paper is to strongly encourage further efforts
to observe OGLE-2000-BUL-43, as this may lead the first complete determination of
the lens parameters. We could even consider a confirmation of the
predictions from Figure~6 to be an ultimate proof of our
understanding of the microlensing geometry. This is particularly
important since the lens model may not be unique. For example, we found
another model (see Table 1, model P$'$) that has $\chisq=320.8$ but with the blending parameter
$f=0.77$. This model predicts a much lower peak ($I_{\rm peak}=12.2$) for an
observer close to Jupiter. Even late space observations will be useful
for distinguishing these two models. For example, the best-fit model predicts 
$I=12.7$ and $I=13.0$ on April 1 and May 1, 2001 respectively, 
while the slightly worse model predicts $I=13.0$ and
$I=13.2$ on these dates. The difference between these two
models can reach 0.02\,mag in the next season for ground-based
observations and hence may be detectable from the ground as well.
However, the alternative model appears physically unlikely 
since the source star is so bright that one would expect $f$ close to 1, 
as found in our best fit model.  The blending parameter may also be 
constrained by spectroscopic observations (Mao, Lennon \& Reetz 1998). A
high-resolution VLT spectrum has already been taken and is currently
being analyzed (K. Gorski 2000, private communication). It will shed further
light on the stellar parameters (such as surface gravity) and the radial
velocity of the lensed source.

\acknowledgments{We acknowledge Bohdan Paczy\'nski for many inspiring
discussions and comments on the manuscript.
We are indebted to Andrew Gould for a prompt, detailed and insightful
referee's report that substantially improved the paper, particularly
concerning the angular stellar radius estimate. We also thank
Andy Drake for many useful comments. This work was supported by the
grants Polish KBN 2P03D00814 and NSF AST 98-20314.
}


\clearpage

\begin{table}
\begin{center}
\small
\caption{The best standard model (first row) and the best
parallax model with blending (second row) for OGLE-2000-BUL-43. The
third rows shows a parallax fit with slightly worse $\chi^2$ (see \S6).
} 
\vspace{0.3cm}
\begin{tabular}{ccccccccc}
\tableline\tableline
Model & $t_0$ & $\tE$ (day) & $\u0$ & $\Is$
& $\psi$ & $\rEt$ (AU) & f & $\chisq$  \\
\tableline
S & $1898.7\pm 0.1$ & $169.6\pm 0.3$ & $0.0\pm 0.002$ & $13.5366\pm 0.0004$
& --- & --- & --- & 9025.2 \\
P & $1893.4\pm 1.0$ & $156.4\pm 4.4$ & $0.27\pm 0.01$ & $13.5406 \pm
0.0004$& $3.024 \pm 0.005$ & $3.62\pm 0.18$ & $0.911\pm 0.056$ & 314 \\
P$^\prime$ & $1842.5\pm 0.9$ & $158.2\pm 4.2$ & $-0.11\pm 0.01$ & $13.5406 \pm
0.0004$ & $3.017 \pm 0.007$ & $4.79\pm 0.22$ & $0.77\pm 0.04$ & 320.8 \\
\tableline
\end{tabular}
\end{center}
\end{table}
\normalsize 

\clearpage

\begin{figure}[t]
\plotfiddle{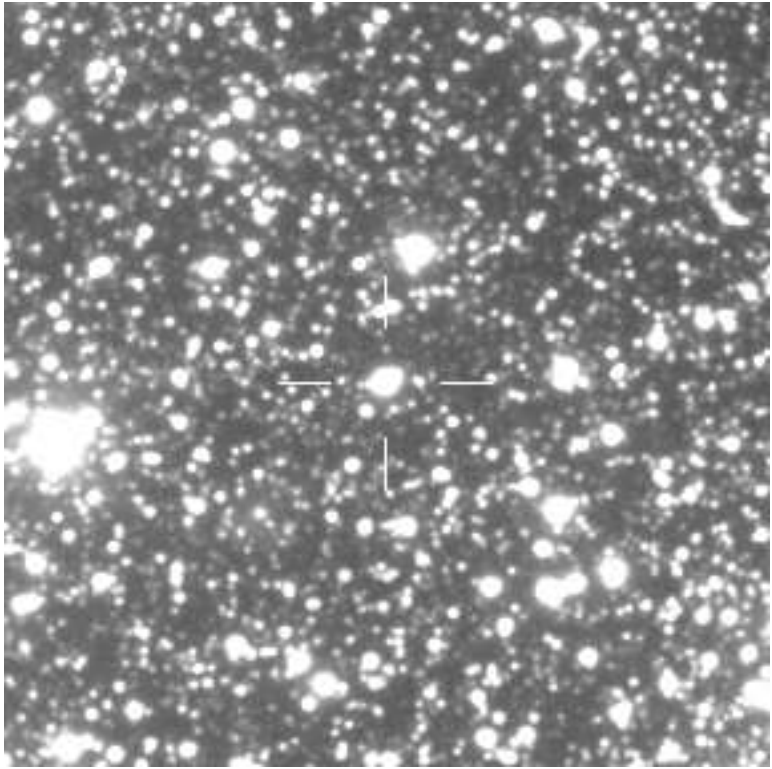}{10cm}{0}{100}{100}{-150}{0}
\caption{Finding chart for the OGLE-2000-BUL-43 microlensing event. The size
of $I$-band subframe is ${120\arcsec\times120}\arcsec$; North is up and
East to the left.}
\label{fig:curves}
\end{figure}

\begin{figure}
\plotone{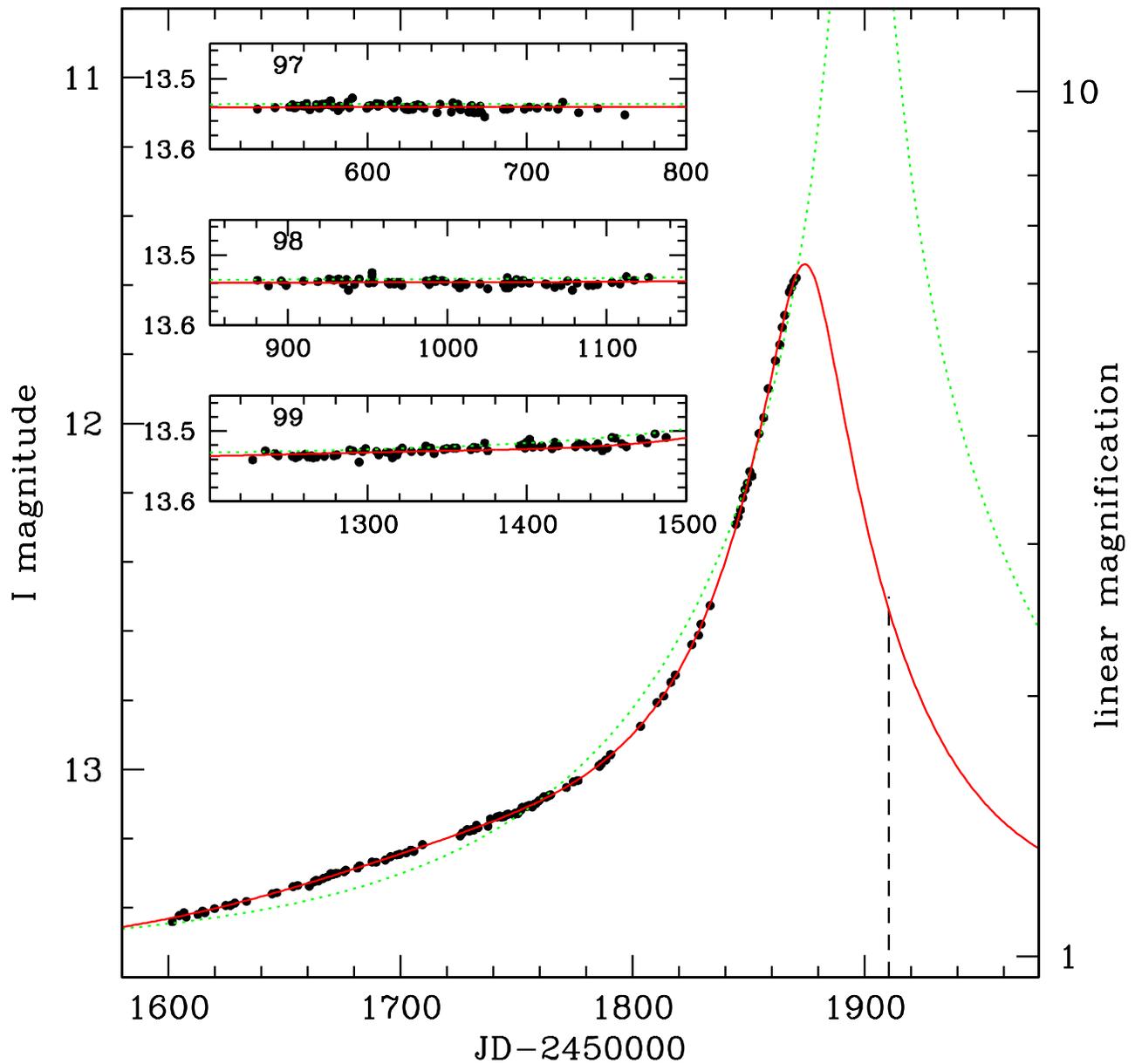}
\caption{$I$-band light curve for the microlensing event
OGLE-2000-BUL-43. The magnitude scale is shown on the left $y$-axis,
while linear magnification is shown on the right $y$-axis.
The dotted line is the standard model while the solid line is the
best-fit model that takes into account the parallax effect and blending
(second row in Table 1). The vertical
dashed line marks January 1, 2001, 0UT. The three insets show the
the data points for the 1997, 1998 and 1999 seasons, respectively.}
\end{figure}

\begin{figure}
\plotone{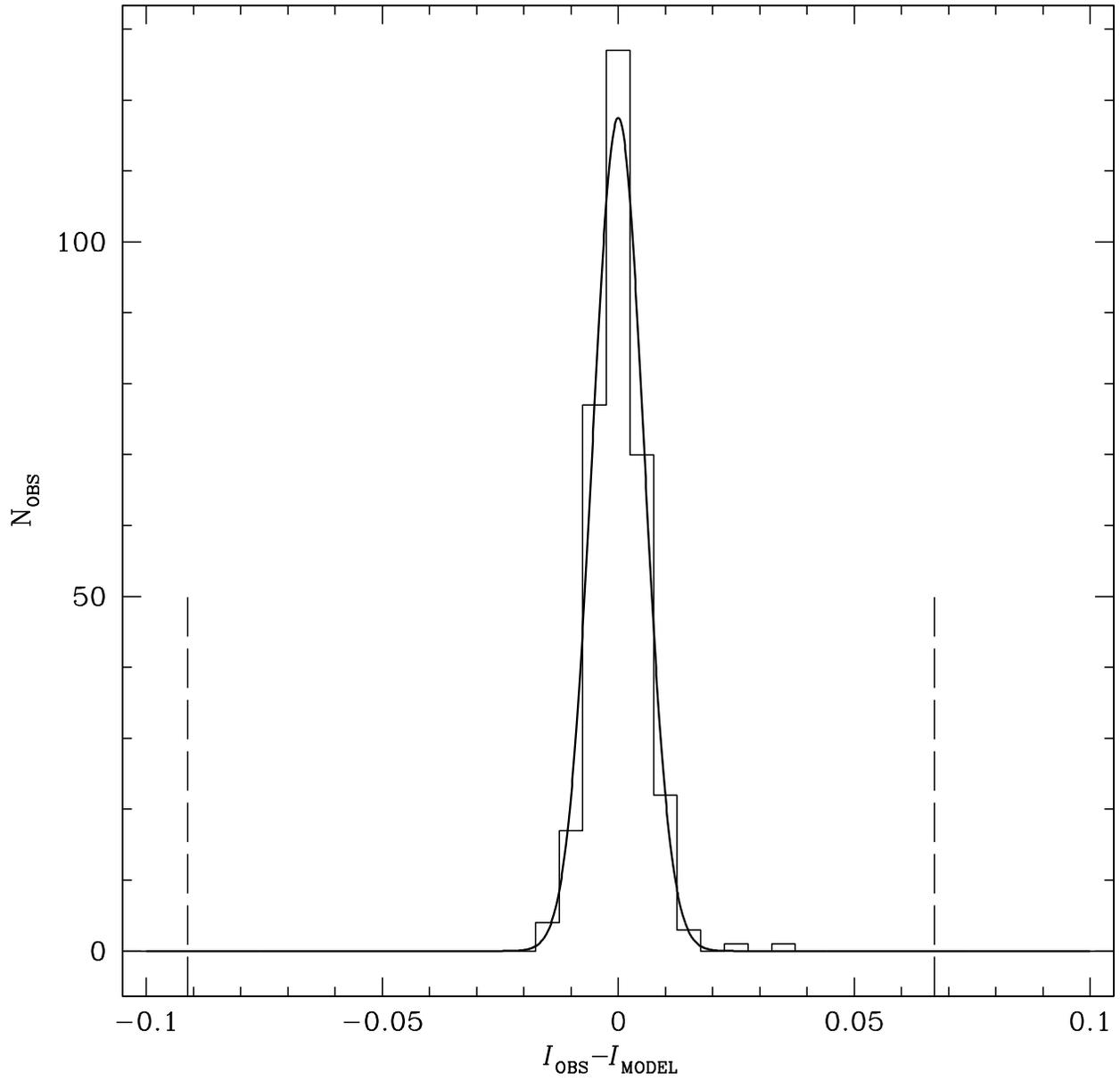}
\caption{Distribution of residuals with respect to the model for
measurements with the DIA pipeline. Width of the bin is 0.005~mag. Sigmas
of fitted Gaussian is 0.0055 mag. Additional
dashed vertical lines indicate the largest differences between the
classical single point microlensing model and the parallax fit.}
\end{figure}

\begin{figure}
\plotone{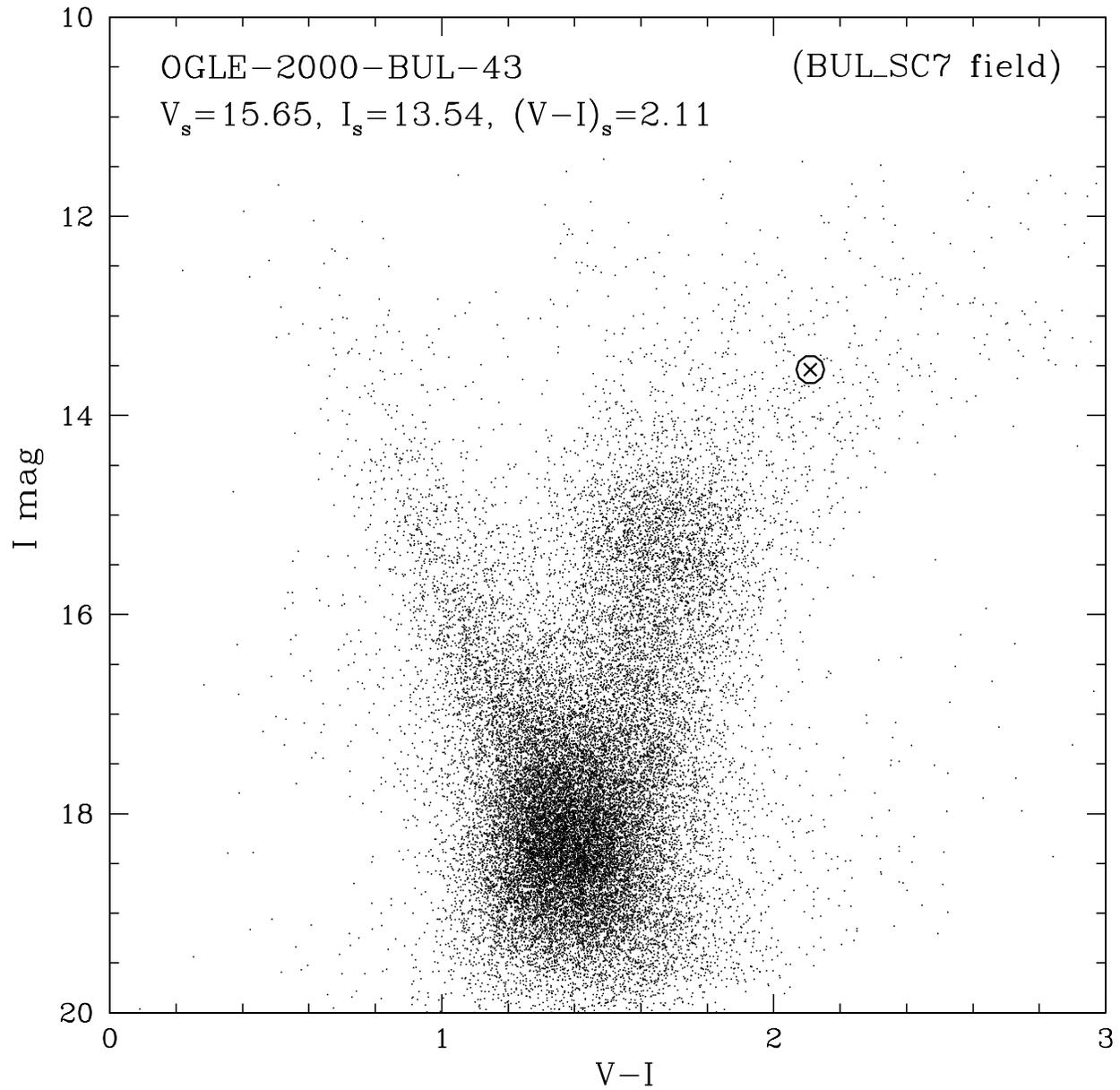}
\caption{Color-magnitude diagram of the BUL\_SC7 field. Only about 10\% of
field stars are plotted by tiny dots. Position of OGLE-2000-BUL-43 event is
marked by cross in the circle.}
\end{figure}

\begin{figure}
\plotone{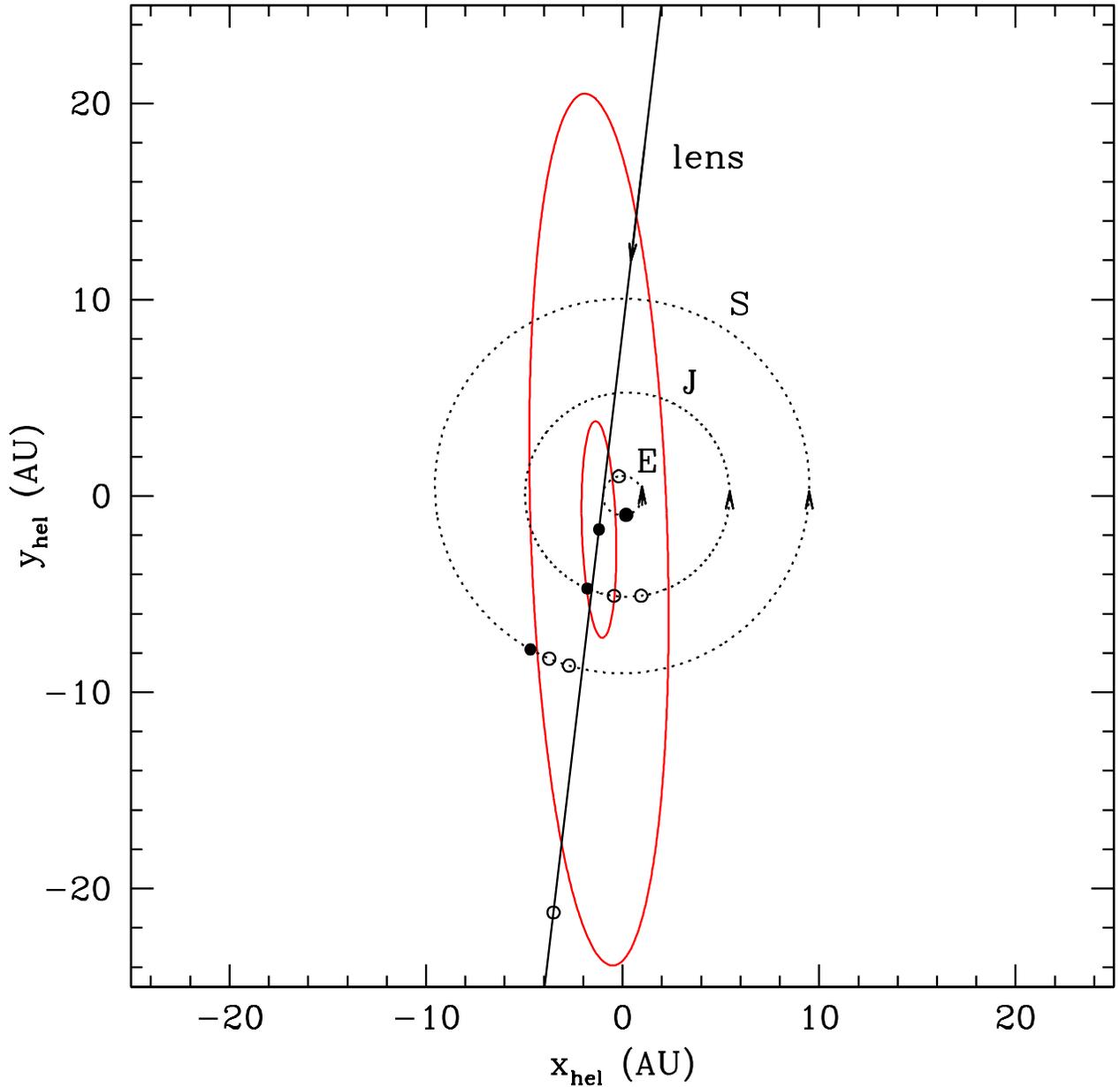}
\caption{Illumination patterns for OGLE-2000-BUL-43 in the heliocentric
ecliptic coordinates on January 1, 2001, 0UT. The $+x$-axis points from the
Sun toward the Earth on the day of Vernal Equinox.
The two solid elliptical curves are the
iso-magnification contours with magnification 1.342 and
4, respectively. The three dotted circles are the orbits
of the Earth, Jupiter and Saturn,
respectively. The solid filled dots on the Earth, Jupiter and Saturn
orbits indicate their positions on January 1, 2001, while the open dots
indicate their positions every half a year in the future. The straight line 
indicates the lens trajectory and the dot symbols have the same
meaning as those on the planetary orbits.
The directions of motions are indicated by arrows. Notice that the whole
illumination pattern (iso-magnification contours) comoves with the lens.
}
\end{figure}

\begin{figure}
\plotone{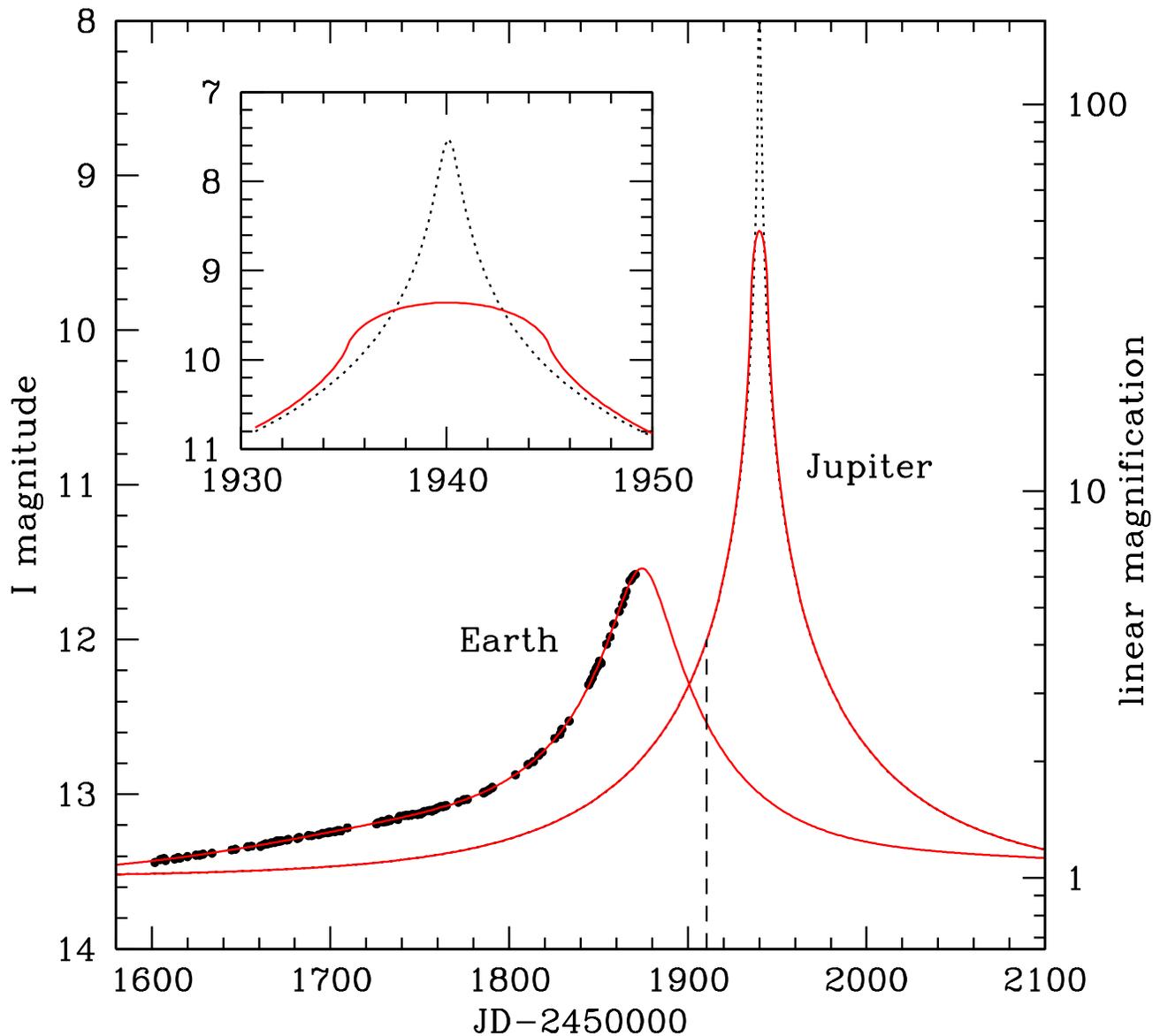}
\caption{Light curve for OGLE-2000-BUL-43 as seen by an observer
close to Jupiter. Notice that it reaches a much higher peak around January 31, 2001 than that on 
the Earth. The vertical dashed line marks January 1, 2001, 0UT
(corresponding to the filled dots in Figure~5).
 The magnitude scale is shown on the left $y$-axis,
while linear magnification is shown on the right $y$-axis.
The dotted line shows the magnification for a point source while
the solid line illustrates the finite source size effect.
The inset shows the light curve close to the peak of the light
curve. 
}
\end{figure}

\begin{figure}
\plotone{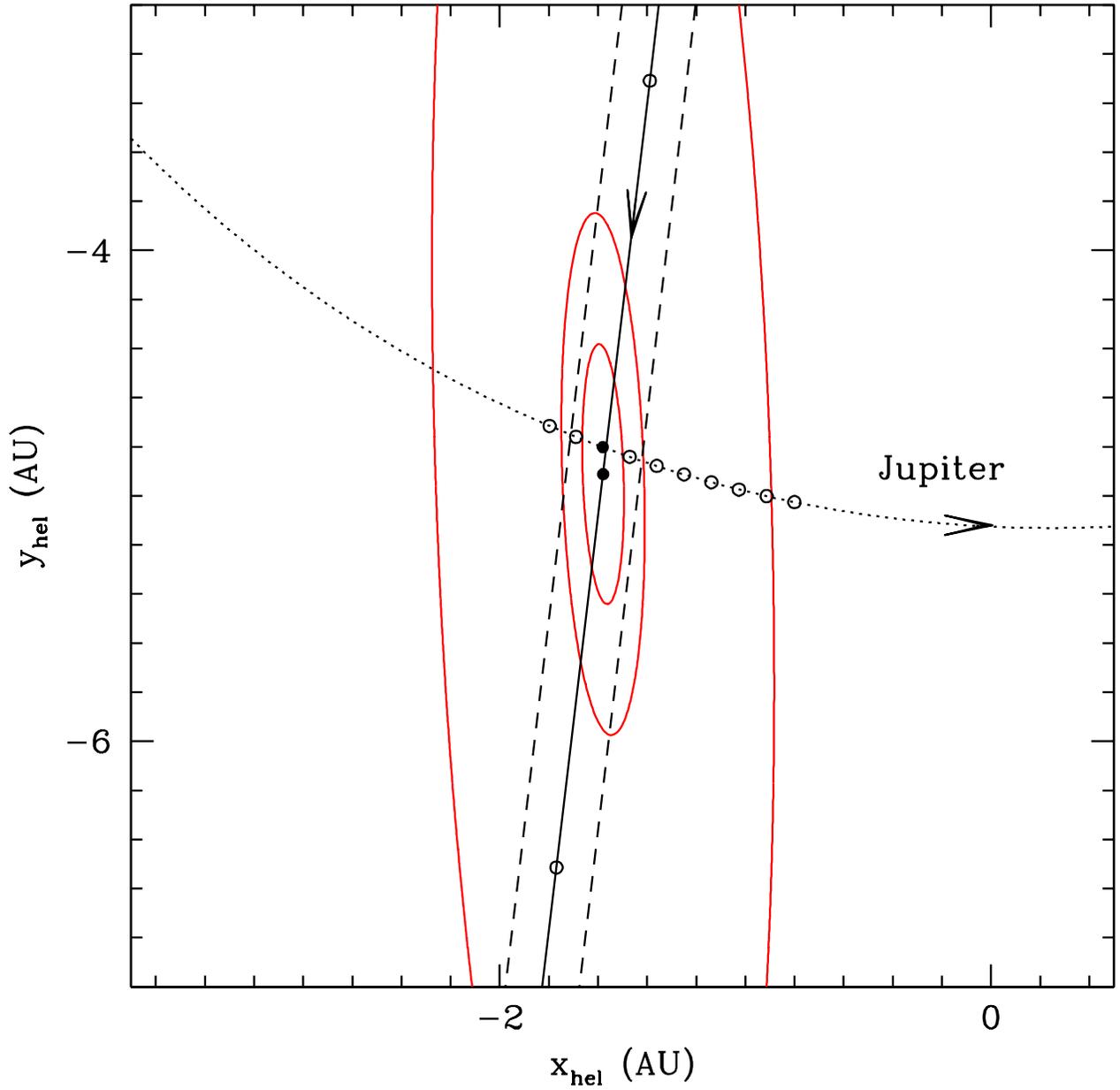}
\caption{Illumination patterns for OGLE-2000-BUL-43 in the heliocentric
ecliptic coordinates on January 31, 2001, 0UT. The notations are
similar to those in Figure~5. The filled dots correspond to $t=1840.5$ while
the open dots are separated by 15 days. The contours
correspond to magnifications of 5, 20 and 40 (from outer to inner),
respectively. The two dashed lines bracket roughly the region that
the finite source size effect can be observed.
}
\end{figure}

\end{document}